\documentclass[apj]{emulateapj}

\newcommand {\lya}    {Ly$\alpha$}   

\newcommand {\HI}     {\ion{H}{1}}   
\newcommand {\OVI}    {\ion{O}{6}}   
\newcommand {\CIII}   {\ion{C}{3}}   

\newcommand {\kms}    {km~s$^{-1}$}
\newcommand {\NOVI}   {$N_{\rm OVI}$}
\newcommand {\NHI}    {$N_{\rm HI}$}

\begin{document}

\title{An O\,VI Baryon Census of the Low-\lowercase{$z$} Warm-Hot Intergalactic Medium}

\author{Charles W. Danforth \& J. Michael Shull\altaffilmark{1}}
\affil{CASA, Department of Astrophysical and Planetary Sciences, University of Colorado, 389-UCB, Boulder, CO 80309; 
danforth@casa.colorado.edu, mshull@casa.colorado.edu}
\altaffiltext{1}{also at JILA, University of Colorado and National Institute of Standards and Technology }

\begin{abstract}
Intergalactic absorbers along lines of sight to distant quasars are a
powerful diagnostic for the evolution and content of the intergalactic
medium (IGM).  In this study, we use the {\it FUSE} satellite to
search 129 known \lya\ absorption systems at $z<0.15$ toward 31 AGN
for corresponding absorption from higher Lyman lines and the important
metal ions \OVI\ and \CIII.  We detect \OVI\ in 40 systems over a
smaller range of column density (log\,\NOVI\ = 13.0--14.35) than seen
in \HI\ (log\,\NHI\ = 13.0--16.0).  The co-existence of \OVI\ and \HI\
suggests a multiphase IGM, with warm photoionized and hot ionized
components.  With improved \OVI\ detection statistics, we find a steep
distribution in \OVI\ column density, $d{\cal N}_{\rm OVI}/dN_{\rm
OVI}\propto N_{\rm OVI}^{-2.2\pm0.1}$, suggesting that numerous
weak \OVI\ absorbers contain baryonic mass comparable to the rare
strong absorbers.  Down to 30 m\AA\ equivalent width (\OVI\ $\lambda
1032$) we find an absorber frequency $d{\cal N}_{\rm OVI}/dz \approx
17 \pm 3$.  The total cosmological mass fraction in this hot gas is at
least $\Omega_{\rm WHIM} = (0.0022\pm0.0003) [h_{70} (Z_{\rm O}/0.1
Z_{\odot})(f_{\rm OVI}/0.2)]^{-1}$ where we have scaled to fiducial
values of oxygen metallicity, \OVI\ ionization fraction, and Hubble 
constant.  Gas in the WHIM at $10^{5-6}$~K contributes at least 
$4.8 \pm 0.9$\% of the total baryonic mass at $z < 0.15$.   
We then combine empirical scaling relations for the observed 
``multiphase ratio", \NHI/\NOVI\ $\propto$ \NHI$^{0.9 \pm 0.1}$, and 
for hydrogen overdensity in cosmological simulations,  
\NHI\ $\propto \delta_H^{0.7}$, with the H~I photoionization correction 
to derive the mean oxygen metallicity, 
$Z_{\rm O} \approx (0.09 Z_{\odot})(f_{\rm OVI}/0.2)^{-1}$ in the 
low-$z$ multiphase gas.  Given the spread in the empirical relations
and in $f_{\rm OVI}$, the baryon content in the \OVI\ WHIM could 
be as large as 10\%.  Our survey is based on a large improvement in the 
number of \OVI\ absorbers (40 vs.\ 10) and total redshift pathlength 
($\Delta z\approx2.2$ vs.\ $\Delta z\approx0.5$) compared to earlier surveys.
\end{abstract}



\section{Introduction}

One of the great, unanticipated legacies of the {\it Far Ultraviolet
Spectroscopic Explorer} (FUSE) mission is surely the detection of
\OVI\ absorption lines along extragalactic sight lines from the
warm-hot intergalactic medium (WHIM) at $T=10^{5-7}$~K, as predicted
by cosmological simulations \citep{CenOstriker99,Dave01}.  Hot gas in
the intergalactic medium (IGM) is produced by shocks generated by
gravitational instability during the formation of large-scale
structure, and its detection in highly ionized oxygen is an indicator
of widespread metal transport into the IGM through feedback from
galaxy formation.  The \OVI\ absorption probes gas at $10^{5-6}$~K,
somewhat cooler than the bulk of the WHIM, which has been shock-heated
to temperatures of several $\times10^6$~K.  This latter, very hot gas
is only detectable through weak X-ray absorption lines from higher
ions (e.g., \ion{O}{7}, \ion{O}{8}, \ion{Ne}{8}, \ion{N}{6},
\ion{N}{7}) as described recently by \citet{Nicastro05}.

The search for the WHIM gas has now begun in earnest, using sensitive
UV resonance absorption lines, primarily the \OVI\ doublet (1031.926,
1037.617~\AA) which is $\sim$100 times more sensitive than the X-ray
transitions of \ion{O}{7} (21.602~\AA) or \ion{O}{8}
(18.97~\AA). Thus, the initial WHIM detections of \OVI\ were made by
{\it FUSE} at $z\leq0.15$ and by the {\it Hubble Space Telescope}
(HST) at $z\geq0.12$, but over a modest total redshift pathlength
$\Delta z\approx0.5$ \citep{Tripp00,Savage02}.  For the portion of
WHIM containing \OVI, the estimated baryon fractions were $\sim$5\%
for O/H metallicity equal to 10\% of the former solar value,
$7.41\times10^{-4}$ \citep{Grevesse96}. Because the solar oxygen
abundance has recently been re-measured at 
(O/H)$_{\odot} = 4.90 \times 10^{-4}$ \citep{abundanceref}, we have
re-scaled our baryon estimates to this lower value.  These estimates
have large uncertainties, including untested assumptions of uniform
metallicity, \OVI\ ionization equilibrium, and multiphase IGM
structure.

The initial \OVI\ searches reported only a handful of absorbers --
four systems by \citet{Tripp00} and six by \citet{Savage02}.  At low
redshift, down to 50~m\AA\ equivalent width in \OVI\ $\lambda1032$,
the number of \OVI\ absorbers per unit redshift is 10--15\% of that
found in the \HI\ \lya\ surveys: $d{\cal N}_{\rm OVI}/dz \approx
14^{+9}_{-6}$ \citep{Savage02} versus $d{\cal N}_{\rm HI}/dz \approx
112 \pm 9$ \citep{Penton4}.  In this study, we dramatically increase
the total surveyed pathlength to $\Delta z>2$ and the total number of
\OVI\ absorbers to 40.  We also study the relationship between
absorbers from the WHIM and the warm neutral medium (WNM;
$10^{3.5-4.5}$~K).  With the increased number of absorbers, we can
begin to look at WHIM absorption in a statistical manner, particularly
the distribution in \OVI\ column density.

\begin{figure*}
  \epsscale{1}\plottwo{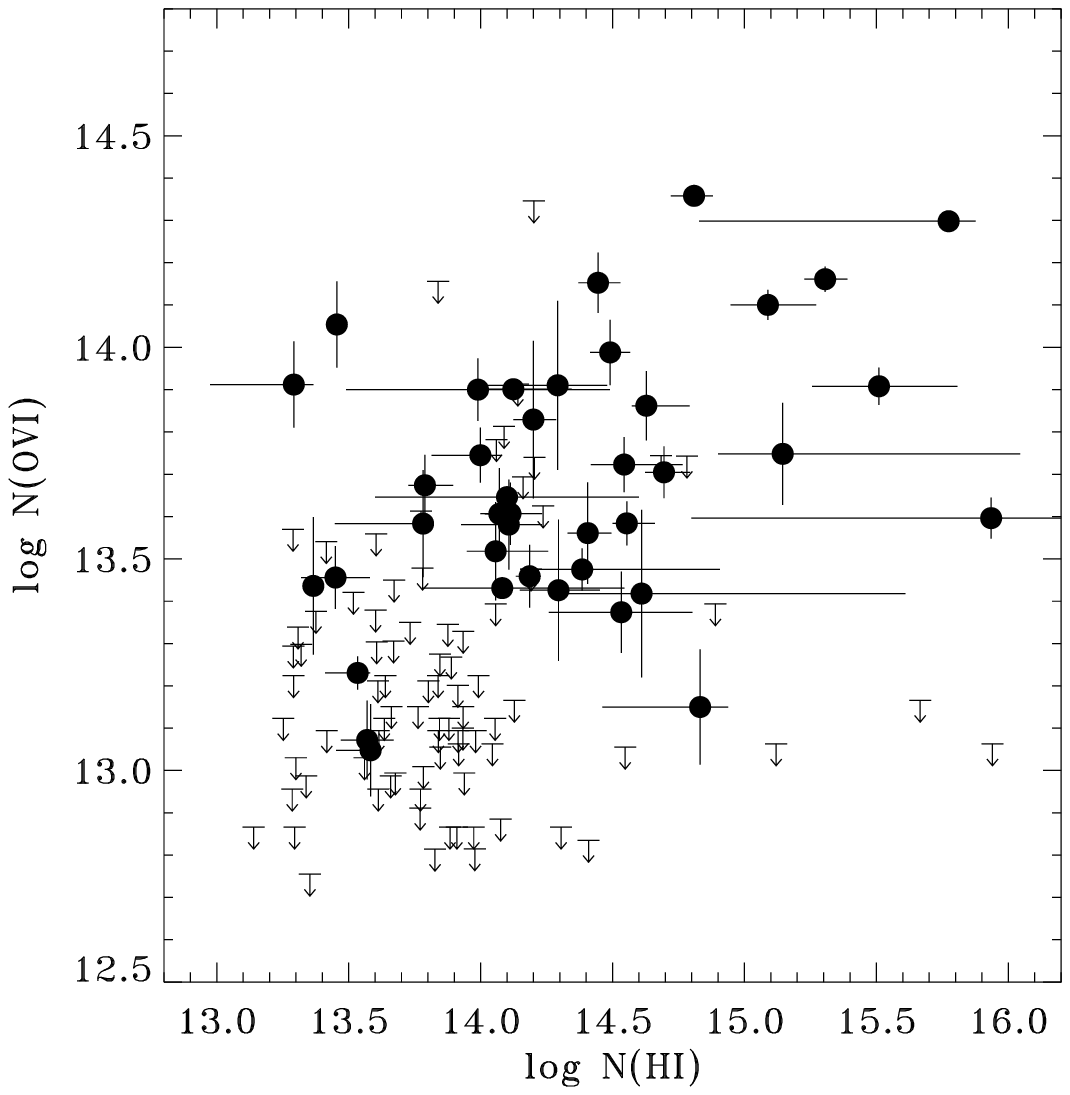}{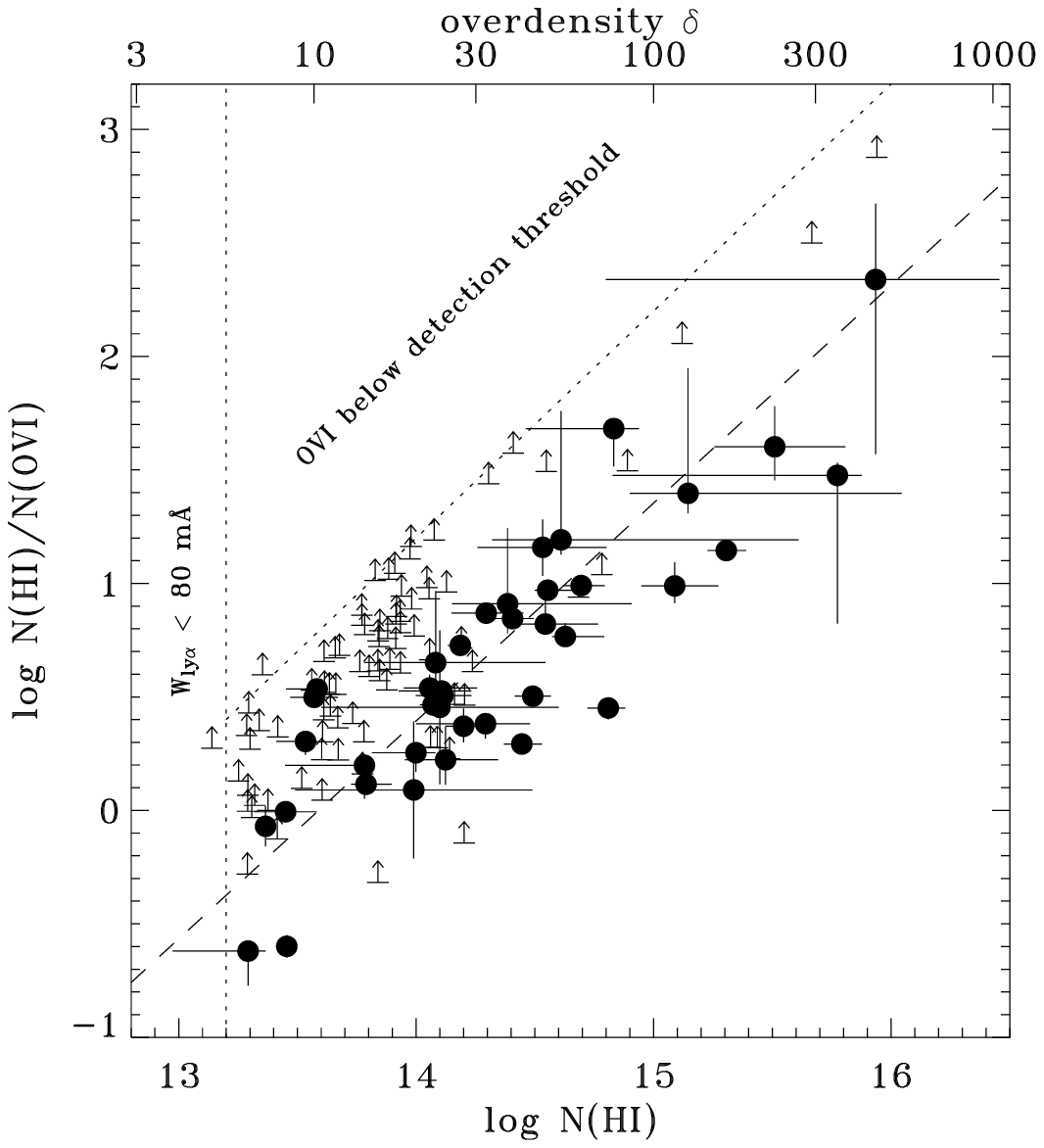}
  \caption{Comparison of H\,I and O\,VI column densities in 124 known
  \lya\ absorbers.  The left panel shows a mild correlation, if any,
  between \NHI\ and \NOVI.  While \NHI\ varies over a factor of nearly
  1000, \NOVI\ varies by a factor of only 20.  The right panel shows
  the ``multiphase ratio'', \NHI/\NOVI, with a dispersion of less than
  one dex over nearly three orders of magnitude in \NHI, suggesting
  that photoionized H~I and warm-hot gas occupy different phases of the 
  IGM.  The multiphase ratio is well fitted by a power law, 
  \NHI/\NOVI\ $= 2.5 \times 
  [{\rm N}_{\rm HI}/10^{14}~{\rm cm}^{-2}]^{0.9 \pm 0.1}$ 
  (dashed line).  \NHI\ can be related to the overdensity, 
  $\delta_H \equiv \rho/ \langle\rho\rangle$, as discussed in \S~4 and 
  shown along the top axis.}
\end{figure*}

\section{Observations and Data Analysis}

We began our survey with published lists of intervening \lya\
absorption systems toward low-$z$ AGN obtained from GHRS and STIS
surveys by \citet{Penton1,Penton4}.  Other sight lines were covered by
literature sources or measured at the University of Colorado from
STIS/E140M spectra as discussed in the upcoming Danforth, Shull, \&
Rosenberg (2005, hereafter Paper~II).  We disregarded any weak \lya\
absorbers (W$_\lambda<80$~m\AA, log\,$N_{\rm HI}\la13.2$) and searched
the {\it FUSE} data for higher-order Lyman lines plus \OVI\ and \CIII\
counterparts. We have been conservative in our identification of
\OVI\ systems, requiring unambiguous ($\geq 4\sigma$) features, and
examining multiple channels of {\it FUSE} data and both lines of the
\OVI\ doublet when possible.  In all, we analyzed 171 absorbers in 31
{\it FUSE} sight lines.  We measured all available Lyman series lines
to determine accurate \NHI\ and doppler $b$ values for \HI\ via
curve-of-growth concordance curves \citep{Shull00}.

Of the \lya\ absorbers, 129 were at $z\leq0.15$, where \OVI\
absorption could potentially be observed.  We detected \OVI\ at
$4\sigma$ or greater level in one or both lines of the doublet for 40
absorbers, and we obtained $4\sigma$ or greater upper limits in 84
other cases.  The remaining five absorbers fell on top of airglow or
strong lines from the interstellar medium (ISM), were blended, or were
in some other way inaccessible.  


We determined metal-ion column densities via Voigt profile fits and/or
the apparent column method \citep{SavageSembach91}. We assumed that
any saturation in these lines was mild, and that profile fits
accurately determine the column density.  In cases where we detected
both \OVI\ lines, we assigned a weighted mean of the column densities
to \NOVI.  We describe the full details of absorber selection and data
analysis in Paper~II.

\section{The Multiphase IGM}

The \OVI\ lines are ideal tracers (at $10^{5.5\pm0.3}$ K) for the
warm-hot ionized medium (WHIM; $10^5-10^7$~K), while \HI\ traces the
photoionized warm neutral medium (WNM; $10^{3.5-4.5}$~K).  The left
panel of Figure~1 shows a weak correlation, if any, between \NOVI\ and
\NHI.  While \NHI\ varies over nearly three orders of magnitude,
\NOVI\ is detected between $10^{13}$ cm$^{-2}$ and a few times
$10^{14}$ cm$^{-2}$ (a factor of $\sim$20).  To gauge the relative
amounts of warm photoionized gas (\HI) and hot collisionally ionized
gas (\OVI), we use the ``multiphase ratio'', \NHI/\NOVI. This ratio
was defined previously \citep{Shull03} as a means of assessing the
range in contributions from photoionized gas (\HI) and collisionally
ionized gas (\OVI), which often appear to be associated kinematically
\citep{Sembach03,Collins04}.

As shown in the right panel of Figure~1, the \NHI/\NOVI\ ratio
exhibits a strong correlation with \NHI, with a typical range
\NHI/\NOVI\ $\approx$ 0.5--50 and a dispersion of less than one (dex).
Higher \HI\ column absorbers typically display a higher multiphase
ratio, while weak \HI\ absorbers have $N_{\rm OVI} \approx
(0.1-2)N_{\rm HI}$.  We are well aware of the fact that plots such as
Figure~1 are subject to the effects of correlated errors in \NHI.
However, this ratio has physical utility in deriving a mean (O/H)
metallicity in multiphase gas (see \S~4), when combined with the empirical
correlation between \NHI\ and hydrogen overdensity, $\delta_H$, seen
in cosmological simulations \citep{Dave99}.

The large range in \NHI\ and the correlation in multiphase ratio
provide evidence that the IGM has at least two phases (WHIM and
WNM). If the \HI\ and \OVI\ materials were well mixed, we would expect
that the multiphase ratio would be constant with \NHI.  We suggest
instead that the WHIM occupies a shell around a warm neutral core.
The outer WHIM shell is heated by a combination of external ionizing
photons (from AGN), shocks from infalling clouds, and possible shocks
from cluster outflows such as superwinds and SNR feedback.  The narrow
range of \NOVI\ compared to \NHI\ implies that the WHIM shell may have
a characteristic column density of $10^{13}-10^{14}$ cm$^{-2}$, while
the neutral core can be arbitrarily large.

In this scenario, the very large values of the multiphase ratio
may arise in ``quiescent gas", possibly located in high-column 
density H~I gas in halos.  The absence of associated \OVI\ suggests 
the lack of shocks at velocities greater than about 150 \kms.  
Many strong H~I (\lya) absorbers have been seen in
proximity (within $200h_{70}^{-1}$ kpc) of bright galaxies 
\citep{Lanzetta95,Chen98,Penton3}.  In our current \OVI\ sample, 
we found three absorbers with extremely high multiphase ratios: very
high neutral hydrogen columns with no detectable WHIM.  In the first
case, the absorber at $cz=1586$ \kms\ toward 3C\,273 shows
log\,$N_{\rm HI}=15.67$ and log\,$N_{\rm OVI}\leq13.17$ and is $71
h_{70}^{-1}$~kpc on the sky from a dwarf ($M_B = -13.9$)
post-starburst galaxy \citep{Stocke04}.  The second case is the
absorber at $cz=23,688$ \kms\ toward PHL\,1811 with log\,$N_{\rm
HI}=15.94$ and log\,$N_{\rm OVI}\leq13.06$.  This system is beyond the
$L^*$ survey depth, so we cannot comment on surrounding galaxies.
Finally, the Lyman-limit system at $cz=24,215$ \kms\ toward PHL\,1811
shows log\,$N_{\rm HI}=18.11$ and log\,$N_{\rm OVI}\leq13.06$.  This
absorber lies $23''$ ($34 h_{70}^{-1}$~kpc) from an $L^*$ galaxy at $z
= 0.0808$ \citep{Jenkins03}.  \citet{Tripp05} note a sub-DLA system at
$z_{\rm abs}=0.00632$ toward PG\,1216+069 with log\,$N_{\rm HI}=19.32$
and log\,$N_{\rm OVI}\leq14.3$ which lies 86 kpc from a sub-L$^*$
galaxy.  These four absorption systems, with multiphase ratios
log\,$[N_{\rm HI}/N_{\rm OVI}]\geq$ 2.5, 2.9, 5.0, and 5.0
respectively, may be shielded from external ionizing flux by adjacent
IGM clouds and have unshocked gas.

Metallicity is one of the great unknowns in interpreting the baryon
content of the WHIM absorbers.  In previous studies, (O/H) has been
assumed to be constant at 10\% solar.  However, one might expect that
metallicity variations could play a role in the multiphase ratio:
outflows of material from galaxies should be denser and more enriched
than primordial IGM clouds.  This would contribute a negative slope in
the multiphase ratio plot (Figure~1b).  Instead, the denser
(high-\NHI) absorbers show a relatively lower ratio of \OVI/\HI,
analogous to lower metallicity.  While metallicity variations almost
certainly exist in the IGM, they are not the main cause of the
positive slope in the multiphase ratio plot.  Some of this effect
could arise from changes in the mean \OVI\ ionization fraction,
typically chosen to be the maximum value, $f_{\rm OVI}\approx0.2$ at
$T_{\rm max}=10^{5.45}$~K in collisional ionization equilibrium
\citep{SutherlandDopita93}.  This fraction is expected to vary,
depending on the range of shock velocities that produce the \OVI\
\citep{Heckman02,Rajan05}.


\section{The Hot Baryon Content of the Universe}

\begin{figure}
  \epsscale{1.2}\plotone{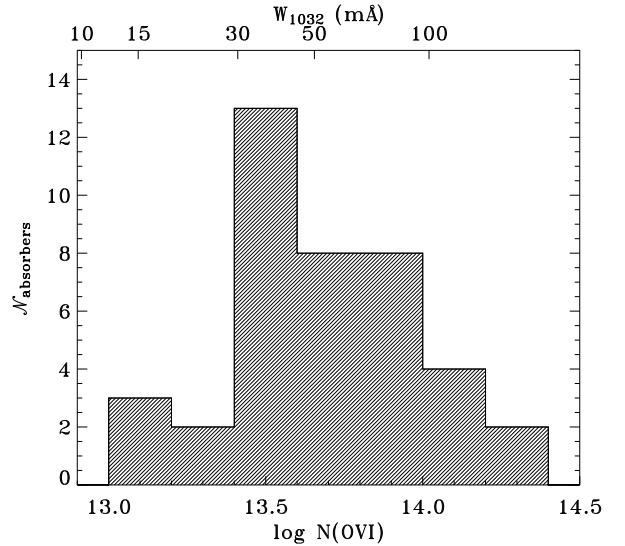}  
  \caption{Column density histogram for the observed O\,VI absorbers.
  Approximate equivalent widths for O\,VI $\lambda$1032 are shown
  along the top axis.  The decreasing number of absorbers at low
  column densities suggests that our survey is incomplete for weaker
  absorption lines.}
\end{figure}

Using \OVI\ as a tracer of the $10^{5-6}$~K portion of the WHIM, we
can employ the number of absorbers per unit redshift, $d{\cal N}_{\rm
OVI}/dz$, to determine $\Omega_{\rm WHIM}$, the fraction of the
critical density contributed by this WHIM gas.  Our detection
statistics are shown in Figure~2 for \OVI.

The total redshift pathlength of our survey is a function of
equivalent width, and our survey is more complete for strong absorbers
than for weak absorbers.  The equivalent-width sensitivity,
$W\rm_{min}(\lambda)$, is a function of spectrograph resolution,
$R=\lambda/\Delta\lambda$, and the signal-to-noise ratio (S/N) of the
data per resolution element.  Thus, for a $4\sigma$ detection limit,
we define $W_{\rm min}(\lambda)=4(\lambda/R)/(S/N)$.  We assume
$R=20,000$ and calculate $W_{\rm min}$ profiles for each dataset based
on S/N measured every 10 \AA.  Strong instrumental features, ISM
absorption, and airglow lines are masked out (S/N set to zero).  From
this procedure, we calculate $N_{\rm min}(\lambda)$ for both \OVI\
transitions, using curves of growth with $b_{\rm OVI} = 25$ \kms.  By moving
an absorber doublet along the profile as a function of $z_{\rm abs}$,
we generate a profile $N_{\rm min}(z)$.  By adding up the total path
length for each absorber at each $N_{\rm min}(z)$, we determine the
relationship between \NOVI\ and $\Delta z$ as shown in Figure~3.  With
a total high-$N_{\rm OVI}$ pathlength $\Delta z=2.21$, we are at least
80\% complete in \OVI\ detection down to log\,\NOVI\ = 13.4
($W_\lambda=30$~m\AA\ in the 1032 \AA\ line).  However, $\Delta z$ and
the completeness fall off rapidly for weaker absorbers.  Dividing
Figure~2 by Figure~3, we obtain the profile of $d{\cal N}_{\rm
OVI}/dz$ as a function of $N_{\rm OVI}$ (Figure~4).  To first order,
this procedure should correct for incompleteness in our \OVI\ survey.

\begin{figure}
  \epsscale{1.2}\plotone{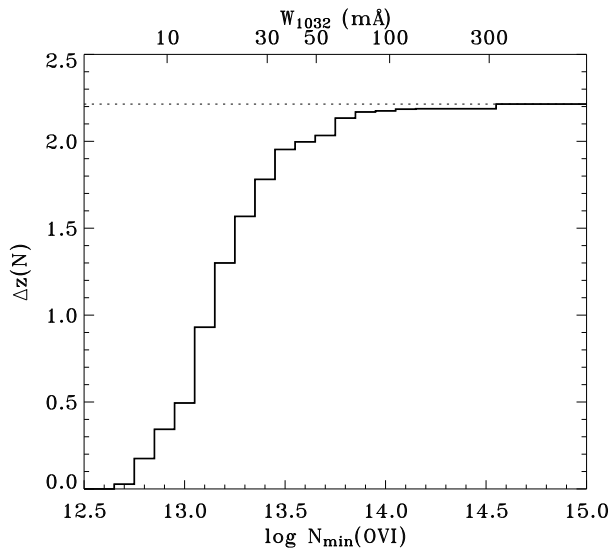} 
  \caption{Cumulative {\it FUSE} redshift pathlength, $\Delta z$, as
  function of 4$\sigma$ absorption-line detection limits in O\,VI
  column density and $\lambda1032$ equivalent width.  Strong lines can
  be detected in poor-quality data. Weak lines can only be detected in
  the highest-S/N observations and thus contribute less total path
  length.  We use the $\Delta z$ profile to correct for incompleteness
  in our O\,VI survey.}
\end{figure}

\begin{figure}[b]
  \epsscale{1.2}\plotone{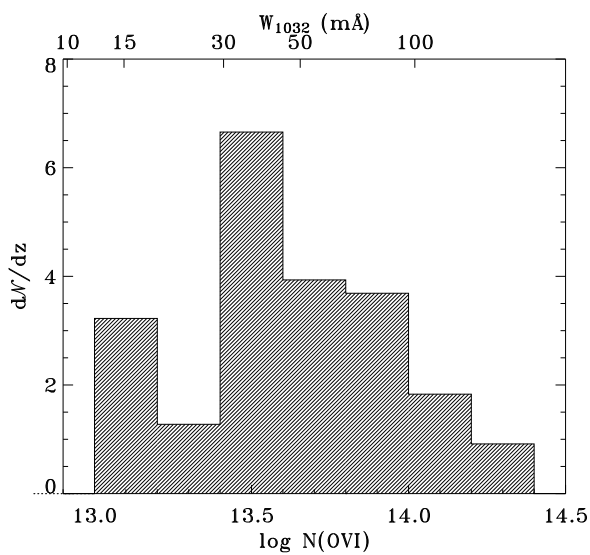} 
  \caption{Number of O\,VI absorbers per column density bin per unit
  redshift after our first-order completeness correction.  Even after
  this correction, the number of O\,VI absorbers appears to fall off
  rapidly below log\,\NOVI $\leq 13.4$.  However, the statistics are
  poor with only five absorbers in the lowest two bins.}
\end{figure}

Our incompleteness-corrected value of the absorber frequency, $d{\cal
N}_{\rm OVI}/dz$, compares well with previous, uncorrected values
(Table~1).  For \OVI\ absorbers with $W_{\lambda} \geq 50$~m\AA\ (in
$\lambda 1032$), we find $d{\cal N}_{\rm OVI}/dz = 9\pm2$, somewhat
lower than the value, $d{\cal N}_{\rm OVI}/dz=14^{+9}_{-6}$, found by
\citet{Savage02} using six \OVI\ absorbers toward PG\,0953+415.
For the weaker \OVI\ absorbers, we find $d{\cal N}_{\rm OVI}/dz =
17\pm3$ ($\rm W_\lambda\geq30$~m\AA) and $d{\cal N}_{\rm OVI}/dz =
19\pm3$ ($\rm W_\lambda\geq15$~m\AA) from our total sample of
$40^{+7}_{-6}$ measured \OVI\ absorbers.  Uncertainties are based on
single-sided $1\sigma$ confidence limits in Poisson statistics
\citep{Gehrels86}.  \citet{Tripp00} found $d{\cal N}_{\rm OVI}/dz>17$
at 90\% confidence for $\rm W_\lambda\geq30$~m\AA, based on four
absorbers toward H1821+643.

A recent study \citep{Tripp04} using STIS/E140M data finds $d{\cal
N}_{\rm OVI}/dz=23\pm4$ for $\rm W_\lambda\geq30$~m\AA\ based on 44
\OVI\ absorbers.  This sample was taken at slightly higher redshift
($0.12\leq z_{\rm abs}\leq0.57$) than our sample ($z_{\rm abs}\leq
0.15$) and shows a higher value of $d{\cal N}_{\rm OVI}/dz$.  This is
not predicted by simulations, which predict an {\it increasing} WHIM
fraction at recent epochs.  However, the difference in redshift
between the two samples is small, and the discrepancy in $d{\cal
N}_{\rm OVI}/dz$ may be a result of cosmic variance.

\begin{figure}
  \epsscale{1.2}\plotone{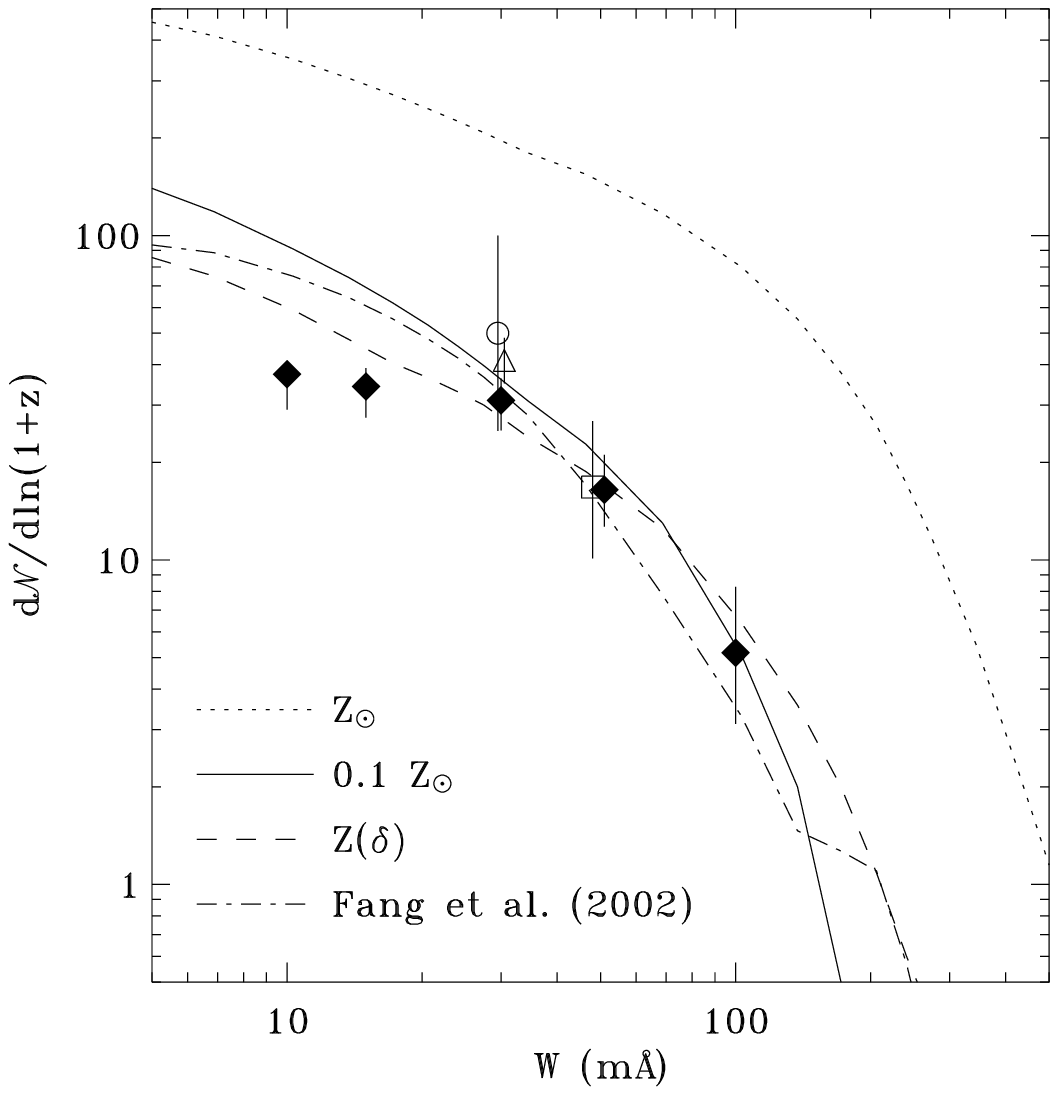}
  \caption{Cumulative equivalent width distribution for O\,VI
  absorbers after Figure~10 of \citet{Chen03}.  Our data points
  (filled diamonds) at (O\,VI $\lambda1032$) equivalent width detection 
  thresholds of 10, 15, 30, 50, and 100 m\AA\ are slightly lower than 
  values from \citet{Savage02} (square), \citet{Tripp00} (circle), and
  \citet{Tripp04} (triangle).  Models from \citet{Chen03} show the
  simulated distribution at a range of metallicities: $Z=0.1Z_\odot$
  (solid curve), $Z=Z_\odot$ (dotted), and an overdensity-dependent
  metallicity $Z=Z(\delta_{\rm H})$ (dashed).  The $Z=0.5\,Z_\odot$ model
  used by \citet{Fang02} (dot-dashed) represents a more simplified 
  physical model.}
\end{figure}

Recent cosmological simulations of the X-ray forest predict a
distribution of \OVI\ absorbers and provide a convenient way to check
current simulations with observed data.  \citet{Chen03} model the
X-ray forest assuming a $\Lambda$CDM model similar to \citet{Dave01},
both collisional ionization and photoionization from a UV background,
radiative cooling, and a range of different metallicities.  The cumulative
distribution of \OVI\ absorbers ($d{\cal N}_{\rm OVI}/d{\rm ln}[1+z]$)
down to a minimum equivalent width is drawn from these models 
and provides a convenient comparison with observed statistics 
from our work and previous surveys (see Figure~5). 

The simulation with $Z=0.1\,Z_\odot$ shows a reasonably good match to
the observed cumulative distribution of \OVI\ absorbers in the local
universe.  Weaker absorbers ($W_\lambda<30$ m\AA) are overpredicted
compared with observations, but this may be a matter of small-number
statistics or differing detection thresholds between observations and
simulations.  A simulation in which metallicity is a function of
overdensity fits the data slightly better than the fixed-metallicity
model.  However, \citet{Chen03} caution that there is a substantial
scatter in the simulations, so that the lower-metallicity curves are
essentially indistinguishable.  The $Z=0.5\,Z_\odot$ model of
\citet{Fang02} also fits the observed data reasonably well, but it is
based on a less physical simulation with no photoionization and no
radiative cooling.  The solar metallicity model of the IGM can clearly
be ruled out by our observations.  Simulations at lower metallicities
($Z = 0.01\,Z_\odot$) would be helpful, given the wealth of new
observational results.

\begin{deluxetable}{lcccl}
\tabletypesize{\footnotesize}
\tablecolumns{5} 
\tablewidth{0pt} 
\tablecaption{IGM O\,VI Absorber Statistics}
\tablehead{\colhead{Criteria}        &
           \colhead{${\cal N}_{\rm abs}$}  &
	   \colhead{$d{\cal N}/dz$}  &
           \colhead{$\Omega_{\rm WHIM}$\tablenotemark{a}} &
	   \colhead{reference} }
\startdata
 $W\geq10$  m\AA&  40 & $21\pm3$        &  $0.0022\pm0.0003$ & this work \\
 $W\geq15$  m\AA&  38 & $19\pm3$        &  $0.0021\pm0.0003$ & this work \\
 $W\geq30$  m\AA&  35 & $17\pm3$        &  $0.0021\pm0.0004$ & this work \\
 $W\geq50$  m\AA&  19 & $~9\pm2$        &  $0.0016\pm0.0005$ & this work \\
 $W\geq100$ m\AA&   6 & $~3^{+2}_{-1}$  &  $0.0008\pm0.0004$ & this work \\
                &     &                 &                    &            \\ 
 $W\geq30$  m\AA&  44 & $23\pm4$        &  0.0027 & \citet{Tripp04} \\
 $W\geq30$  m\AA&   4 & $>17~~~$        & $>$0.006\tablenotemark{b}& \citet{Tripp00} \\
 $W\geq50$  m\AA&   6 & $14^{+9}_{-6}$~~&$\geq$0.003~~~& \citet{Savage02} 
\enddata     
\tablenotetext{a}{All $\Omega_{\rm WHIM}$ values have been converted to a consistent set of assumptions:  $f_{\rm OVI}=0.2$, $Z=0.1\,Z_\odot$, $H_0=70h_{70}~{\rm km~s^{-1}~Mpc^{-1}}$, and (O/H)$_\odot=4.9\times10^{-4}$ \citep{abundanceref}.}
\tablenotetext{b}{All four O\,VI absorbers in \citet{Tripp00} lie along the unusually rich H1821+643 sight line.}
\end{deluxetable}

From $d{\cal N}/dz$, we can calculate the contribution to $\Omega_b$
from WHIM gas, as discussed in \citet{Savage02}.  We assume a Hubble
constant $H_0=70 h_{70}$ \kms~Mpc$^{-1}$, rather than their value
$h_{75}$, and we make standard assumptions regarding \OVI\ ionization
fraction and (O/H) metallicity \citep{Tripp00,Savage02}.  We adopt an
ionization fraction, $f_{\rm OVI}=0.2$ characteristic of its maximum
value in collisional ionization equilibrium (CIE) and assume an O/H
abundance 10\% of the solar value.  Of course, CIE is almost certainly
not a valid assumption for the low-density IGM.  As the infalling gas
is shock-heated during structure formation, the ionization states
\OVI, \ion{O}{7}, and \ion{O}{8} undergo transient spikes in
abundance, followed by cooling and recombination.  Recent
time-dependent models of the ionization, recombination, and cooling of
shock-heated, low-density WHIM \citep{Rajan05} find that the mean,
time-averaged \OVI\ ion fraction is $\langle f_{\rm OVI}
\rangle=$8--35\%, over a range of initial temperatures $5.4 \leq
\log T \leq6.2$.  These fractions are compatible (with a factor of two
spread) with the fiducial value, $f_{\rm OVI} = 0.2$, based on the
maximum fraction of \OVI\ at $T_{\rm max} = 10^{5.45}$~K, in CIE
\citep{SutherlandDopita93}.

As shown above (Fig.\ 5), a metallicity $Z=0.1\,Z_\odot$ is reasonably 
consistent with the observed equivalent width distribution of \OVI\ 
absorbers at $z\sim0$. However, using our sample of \OVI\ and \HI\ 
absorbers, we can make a more direct estimate of the (O/H) metallicity.   
To do so, we must estimate the amount of hydrogen associated with
each component of the the multiphase system of photoionized \HI\ and 
collisionally ionized \OVI.  Since the \HI\ and \OVI\ absorbers are 
associated kinematically, we assume they share the same metallicity.  
We then use the empirical relations between \NHI\ and overdensity, 
$\delta_H$, and between the multiphase ratio (\NHI/\NOVI) and \NHI. 

\begin{equation} 
   \left( \frac {N_{\rm HI}} {N_{\rm OVI}} \right) = C_0~N_{14}^{\alpha} \; ,  
\end{equation}
and
\begin{eqnarray}        
   \delta_{\rm H} &\equiv& \frac {n_H} {(1.90 \times 10^{-7}~{\rm cm}^{-3})(1+z)^3}\nonumber \\
                  &\approx& 20~N_{14}^{0.7}~10^{-0.4z}   \; .  
\end{eqnarray}  
The above relations allow us to relate \NHI\ to the physical gas density, 
$n_H$, needed for the hydrogen photoionization correction.
Equation (1) is derived by fitting the multiphase correlation
in Fig.\ 1b over the approximate range $10 \leq \delta_{\rm H} \leq 300$.  
The scaling constant $C_0 = 2.5 \pm 0.2$ is the mean multiphase ratio
at log\,\NHI\ = 14, and the best-fitting slope is $\alpha = 0.9\pm 0.1$.
Equation (2) comes from cosmological simulations \citep{Dave99} 
and relates the hydrogen overdensity, $\delta_H$, to the \HI\ column 
density, N$_{\rm HI} \equiv (10^{14}~{\rm cm}^{-2}) N_{14}$.

From these relations, we can derive a statistical value of the
(O/H) metallicity from the formula, 
\begin{equation}
     \langle N_{\rm O}/ N_{\rm H} \rangle = 
          \langle N_{\rm OVI}/ N_{\rm HI} \rangle       
         \times \left( \frac {f_{\rm HI}} {f_{\rm OVI}} \right) \; .   
\end{equation}
We employ the multiphase ratio, \NHI/\NOVI, together with appropriate ionization 
correction factors, $f_{\rm OVI}$ and $f_{\rm HI}$.  
The \HI\ fraction is derived from photoionization equilibrium in 
the low-$z$ IGM 
\begin{equation}
   f_{\rm HI} = \frac {n_e \alpha_H^{(A)}} {\Gamma_H} =  
      (1.80 \times 10^{-5}) (1+z)^3 \left( \frac {\delta_H}{20} \right) 
      T_4^{-0.726} \Gamma_{-13}^{-1}  \; , 
\end{equation}  
for gas with $n_e = 1.16 n_H$ at temperature $(10^4~{\rm K})T_4$,
photoionized at rate $\Gamma_{\rm H} = (10^{-13}~{\rm s}^{-1})
\Gamma_{-13}$.  Since the mean-free path of a Lyman continuum photon
is very large in the low-overdensity IGM, we use the case-A
recombination rate coefficient $\alpha_H^{(A)} = (4.09 \times
10^{-13}~{\rm cm}^3~{\rm s}^{-1}) T_4^{-0.726}$.  Combining the two
empirical relations with the relation of photoionization equilibrium,
we find that the mean oxygen metallicity of the
\OVI\ absorbers at $\langle z \rangle = 0.06$ is
\begin{equation}
   Z_{\rm O} = (0.09 Z_{\odot}) N_{14}^{-0.2} T_4^{-0.726} \Gamma_{-13}^{-1} 
     \left( \frac {f_{\rm OVI}} {0.2} \right)^{-1}   \; .  
\end{equation}  
Given the uncertainties in these empirical relations, it is remarkable that 
this formula arrives at an oxygen abundance near the fiducial value of 10\%
solar.   In fact, our estimated value, $Z_{\rm O} = 0.09 Z_{\odot}$, is
probably accurate to only a factor of 2.   

\newpage 
We now derive the fractional contribution to closure density of WHIM
baryons associated with hot \OVI,
\begin{eqnarray}   
\Omega_{\rm WHIM}&=&\left(\frac {H_0} {c\,\rho_{\rm cr}}\right) \frac{\mu m_H}{(O/H)_{\odot}~Z~f_{\rm OVI}}\nonumber \\
&&\times\int_{N_{\rm min}}^{N_{\rm max}}\left(\frac{d\cal{N}}{dz}\right)\,\langle N_{\rm OVI}\rangle\,dN_{\rm OVI} \\  
&=&(1.85\times10^{-18})~h^{-1}_{70}~\sum_i\left(\frac{d\cal{N}}{dz}\right)_i\;\langle N_{\rm OVI}\rangle_i\;.\nonumber
\end{eqnarray}
We perform the above sum using the value, $(d{\cal N}/dz)_i$, for each
column-density bin in Figure~4, with $\langle N_{\rm OVI}\rangle_i$
chosen as the mean column density (cm$^{-2}$) in the bin.  We find
$\Omega_{\rm WHIM} = (0.0022\pm0.0003) h_{70}^{-1}$.  Values of
$\Omega_{\rm WHIM}$ for other equivalent width thresholds are listed
in Table~1 along with the results of other studies converted to
uniform values of $H_0$, $f_{\rm OVI}$, and (O/H)$_{\odot}$.

It should be stressed that these estimates of $\Omega_{\rm WHIM}$ are based
on an assumed uniform 10\% solar (O/H) metallicity and that further
uncertainty arises in our assumed \OVI\ ionization fraction.  Changes
in $f_{\rm OVI}$ within the range of mean values \citep{Rajan05} could 
change our result by a factor of two.  Based on the standard
assumptions, our result is slightly lower than the result published by
\citet{Savage02}, $\Omega_{\rm WHIM}\geq0.002 h_{75}^{-1}$.
\citet{Savage02} used an older solar oxygen abundance, (O/H)$_\odot =
7.41\times10^{-4}$, which is 50\% larger than the \citet{abundanceref}
value.  After correcting for this difference, their result for
$\Omega_{\rm WHIM}$ is larger than ours.  \citet{Tripp04} find 
$\Omega_{\rm WHIM}=0.0027$ based on 44 absorbers at $z\geq0.12$.  

Our estimate corresponds to $\Omega_{\rm WHIM}/\Omega_b = 0.048\pm 0.007$, 
so that WHIM gas in the range $10^{5-6}$~K makes up roughly 5\% of the 
baryonic mass in the local universe.  The contribution could exceed 10\% 
if we account for the likely spread in (O/H) metallicities and \OVI\ 
ionization fractions.  Simulations predict that the current universe is 
composed of $\sim30$\% WHIM gas, so our value of $\sim3-10$\% falls 
short of this mark.  However, \OVI\ is only useful as a proxy for gas 
within the lower portion of the WHIM temperature range.  Accounting for 
the hotter $10^{6-7}$~K gas using similar methodology will require
high-sensitivity, high resolution X-ray observations of \ion{O}{7} and
\ion{O}{8} with spectrographs aboard future missions such as Constellation-X 
or XEUS \citep{Fang02,Chen03}.  Limited observational work has been done in
this area \citep{Nicastro05}, but the current generation of X-ray
telescopes are not ideal for this kind of investigation.

The distribution of \HI\ absorbers with column density is often
expressed as a power law with index $\beta$: $d{\cal N}/dN \propto
N^{-\beta}$ \citep{Weymann98,Penton2,Penton4}.  For the first time,
this analysis can be applied to WHIM species.  We find that the
differential number of \OVI\ absorbers with column density is a power
law, $d{\cal N}_{\rm OVI}/dN_{\rm OVI} \propto N_{\rm
OVI}^{-2.2\pm0.1}$ for absorbers with log\,$N_{\rm OVI}\geq13.4$.
This is somewhat steeper than the corresponding \HI\ distribution,
$\beta\sim1.6$ \citep{Penton4} and means that low-column absorbers
contribute comparable amounts to the \OVI\ baryon census as the rare
high-column systems.  Note that in eq.~[6], $\Omega_{\rm WHIM}$ scales
as $N_{\rm min}^{-0.2}$ for $\beta=2.2$.  Even though we correct for
incompleteness in the weaker absorbers, the statistics are still poor
at log $N_{\rm OVI}\leq13.4$.  It is unclear whether the turnover in
$d{\cal N}_{\rm OVI}/dz$ at lower columns is real or a statistical
fluctuation from small numbers of weak absorbers.  Nevertheless, the
weak \OVI\ absorbers appear to make significant contributions to the
baryon mass density.  The steep power law also reinforces our
conclusions about the nature of the multiphase IGM with a core-halo
structure.


We have demonstrated that the contributions from weak \OVI\ absorbers
cannot be neglected in an accurate WHIM baryon census.  The rare
strong \OVI\ absorbers and numerous weak absorbers contribute nearly
equally to $\Omega_{\rm WHIM}$.  Further FUSE observations at high-S/N
will allow us to probe weak \OVI\ absorbers and refine the statistics
at the low-column end of the absorber distribution.  An analysis of
\OVI\ absorbers at higher redshifts ($z>0.5$) would allow us to track
changes in the WHIM density, confirming the decrease in the amount of
shocked gas at higher redshifts predicted by cosmological simulations
\citep{CenOstriker99,Dave99,Dave01}.  

Only by enlarging our sample of multiphase (\HI, \OVI) absorbers
beyond the 40 discussed here will we be able to refine our statistical
estimate for the metallicity of the WHIM.  With a much larger sample
of \OVI\ absorbers, we could use maximum-likelihood techniques to
search for the expected trends: (1) decreasing metallicity at lower
overdensity $\delta_H$ \citep{Gnedin97}; (2) decreasing $\Omega_{\rm
OVI}$ at higher redshift \citep{Dave99}; and (3) trends of \NHI/\NOVI\
with other metal indicators such as \CIII, \ion{C}{4}, or \ion{Si}{4}.

\acknowledgments

We are grateful to Steve Penton, John Stocke, Jessica Rosenberg, Todd
Tripp, Bill Blair, Blair Savage, Ken Sembach, and Jason Tumlinson for 
useful discussions regarding this project.  This work contains data
obtained for the Guaranteed Time Team by the NASA-CNES-CSA {\it FUSE}
mission operated by the Johns Hopkins University, as well as data from
the {\it Hubble Space Telescope}.  Financial support to the University
of Colorado has been provided by NASA/{\it FUSE} contract NAS5-32985
and grant NAG5-13004, by our HST \lya\ survey program 6593, and by
theoretical grants from NASA/LTSA (NAG5-7262) and NSF (AST02-06042).


\end{document}